\documentclass[%
 reprint,
 amsmath,amssymb,
 aps,
floatfix,
]{revtex4-2}
\usepackage{threeparttable} 
\usepackage{multirow}
\usepackage{array}
\usepackage{amsmath,amssymb}
\usepackage{graphicx}
\usepackage{dcolumn}
\usepackage{bm}

\usepackage{hyperref}
\hypersetup{
    colorlinks = true,
    linkcolor = blue,
    citecolor = blue,
    urlcolor = blue,
}
\begin{document}

\preprint{APS/123-QED}

\title{Novel topological subclass in Bardeen-AdS-class black holes}

\author{
  Yu-Die Wan$^{1}$,
  Peng Zhao$^{1}$,
  Zheng-Wen Long$^{1}$\thanks{Corresponding author: zwlong@gzu.edu.cn}
}
\email{zwlong@gzu.edu.cn}

\affiliation{$^{1}$ College of Physics, Guizhou University, Guiyang, Guizhou 550025, People’s Republic of China}

\date{\today}
\begin{abstract}
Based on the $\phi$-mapping topological current theory, we systematically investigate the thermodynamic topology of regular Bardeen–AdS-class black holes coupled to nonlinear electrodynamics, a class of singularity-free geometries whose topological classification remains underexplored. We implement numerical calculations at two typical AdS radii $L=1$ and $L=15$. Within the fundamental $W^{0-}$ topological family, we identify a novel inner secondary subclass $\widetilde{W}^{0-}$. This result only refines and enriches the internal branch structure of the $W^{0-}$ family under the established classification scheme of five topological classes and four topological subclasses for black hole thermodynamics. This new subclass has a vanishing global topological charge $W=0$ with an inner–outer horizon winding-number signature $[-,+]$. Unlike standard $W^{0-}$ solutions that realize topological extension via embedded $(+,-)$ winding pairs, $\widetilde{W}^{0-}$ arises from attaching an extra stable branch to the end of the base solution sequence. Two critical coupling parameters $\hat{m}_{01}$ and $\hat{m}_{02}$ divide the full parameter space into distinct topological phases: Type I black holes with $m_0\ge\hat{m}_{01}$ and Type II solutions with $\hat{m}_{02}<m_0<\hat{m}_{01}$ belong to $\widetilde{W}^{0-}$, while the parameter region $0<m_0\le\hat{m}_{02}$ corresponds to the conventional $W^{0-}$ topology. Our numerical results verify the self-consistency of the thermodynamic topological formalism when applied to singularity-free regular black holes, refine the topological discrimination criteria tailored to nonlinear-electrodynamics regular black holes, and provide theoretical references for subsequent topological investigations of higher-dimensional rotating regular black holes under the same formalism.
\end{abstract}

\maketitle


\section{Introduction}
\label{sec:level1}
Black hole thermodynamics, pioneered by Bekenstein and Hawking\cite{a1,a2}, has emerged as a crucial bridge connecting gravitational theory, statistical physics, and quantum information. The four laws of black hole thermodynamics establish a fundamental correspondence between gravitational systems and conventional thermodynamic systems\cite{a3}. Within the extended phase space formalism, the cosmological constant is interpreted as thermodynamic pressure\cite{a4,a5}, which significantly enriches the phase structure of black holes  and gives rise to abundant phenomena such as van der Waals-like phase transitions\cite{a6}, reentrant phase transitions\cite{a7,a8,a9,a10}, and reversible heat engine cycles\cite{a11,a12}, making black hole thermodynamic chemistry a frontier direction in contemporary gravitational physics\cite{a13,a14,a15}. A large number of studies have shown that the phase behavior of black holes follows universal underlying laws independent of specific models, and the key to revealing such universality lies in adopting a topological perspective beyond equilibrium thermodynamics\cite{a16}.

In recent years, the thermodynamic topological approach has provided a unified framework for characterizing the phase structure of black holes in a universal manner\cite{a17}. In this formalism, black hole solutions are treated as topological defects in the thermodynamic parameter space\cite{a18}. By constructing the topological vector field and evaluating topological invariants, one can achieve the universal classification of phase structures and thermodynamic stabilities without relying on the details of specific gravitational theories or black hole solutions\cite{a19,a20,a21}. This framework has been successfully applied to various singular black holes, including RN, RN-AdS, Schwarzschild, Schwarzschild–AdS\cite{a22},four-dimensional static two-charge AdS, four-dimensional dyonic AdS\cite{a23}, higher-odd-dimensional, multiply rotating Kerr-AdS\cite{a24} and charged static black holes\cite{a25}, classifying them into five fundamental topological classes and four extended subclasses. Extensive topological calculations for a wide spectrum of black hole models strongly demonstrate the universal validity of this topological method for generic black hole spacetimes, revealing the intrinsic and universal characteristics of thermodynamic phase behaviors and stability sequences of black hole systems in both high- and low-temperature limits\cite{a26,a27,a28,a29,a30}.

It is worth noting that the existing topological thermodynamic frameworks have been predominantly constructed for singular black hole solutions with central spacetime singularities. Although topological methods have been preliminarily applied to regular singularity-free black holes\cite{a31,a32}, a systematic and unified exploration of their topological thermodynamic properties and phase structure within a consistent theoretical framework is still lacking. Bardeen-AdS-type spacetimes are typical regular black holes derived from minimally coupled nonlinear electrodynamics, free of central curvature singularities throughout their entire domain. The generalized action framework constructed in Ref. \cite{a33} fully decouples the black hole’s dynamical mass, magnetic charge, and intrinsic coupling parameters of the matter field, enabling a clear separation of the respective contributions of various physical quantities to spacetime geometry and thermodynamic behavior. Self-consistent extended first laws of thermodynamics and the corresponding Smarr scaling identities can be derived within this formalism, resolving the intrinsic thermodynamic inconsistency plaguing the original Bardeen black hole \cite{a34,a35}. The absence of a central curvature singularity constitutes the core geometric feature distinguishing such regular spacetimes from conventional singular black holes including Schwarzschild and Reissner–Nordström black holes. Representative systematic investigations concerning null geodesics and quasinormal modes of regular black holes can be found in Refs. \cite{a36,a37}.

The generalized parameter space divides Bardeen-AdS-class regular black holes into two distinct thermodynamic families, Type I (\(m_0\geq\hat{m}_{01}\)) and Type II (\(0<m_0<\hat{m}_{01}\)), with markedly different phase structures. Type I accommodates three equilibrium black hole states, where the regular Bardeen–AdS-class solution serves as a tunable intermediate configuration that can be stable, metastable, or unstable during phase transitions. By contrast, Type II comprises four black hole phases alongside a horizonless pure Bardeen–AdS-class vacuum, supporting both Hawking–Page-like vacuum-black hole transitions and isothermal equilibria between small and large black holes\cite{a33,a38}. The phase evolution\cite{a39,a40,a41,a42}, stochastic kinetics and phase-space topologies of such singularity-free black holes \cite{a43,a44,a45,a46}deviate drastically from those of conventional singular black holes, motivating three core topological questions investigated in this work: Can the established classification scheme of five topological classes and four topological subclasses fully describe the topological behaviors of regular Bardeen–AdS geometries? Do regular black holes host unreported secondary subclasses confined to a single existing topological family that enrich the internal topological evolution rules of this family? Does the winding-number-based topological formalism remain self-consistent for singularity-free regular spacetimes?

The organization of this paper is outlined as follows. Section \ref{sec:level2} briefly reviews the thermodynamic topological formalism proposed in Ref. \cite{a18}, laying a theoretical foundation for the subsequent comparison with the novel topological subclass identified in Section \ref{sec:level3}. In Section \ref{sec:level3}, we systematically analyze the generic topological categories of Bardeen–AdS-class black holes and introduce the new subclass \(\widetilde{W}^{0-}\), whose distinctive properties are highlighted via detailed comparative analysis. Finally, Section \ref{sec:level4} summarizes our main results and presents prospects for future extensions of this research.

\section{Thermodynamic Topological Approach and Established Classification}
\label{sec:level2}
The topological approach to black hole thermodynamics treats black hole solutions as topological defects in the thermodynamic parameter space and provides a unified classification of black hole thermodynamic phases based on topological invariants\cite{a17,a18,a47}. This method is model-independent and highly universal, built on Duan’s theory of \(\varphi \)-mapping topological currents\cite{a48,a49,a50,a51}. By defining a generalized free energy, constructing a vector field, and computing topological numbers, it establishes a rigorous correspondence between thermodynamic stability and topological characteristics, serving as a core theoretical tool to uncover the intrinsic properties of black hole phase structures\cite{a52,a53}.

Within the framework of finite-cavity thermodynamics, the generalized off-shell Helmholtz free energy of a black hole is defined as\cite{a18,a22,a54}
\begin{equation}
  \mathcal{F} = M - \frac{S}{\tau},  
  \label{eq:1}
\end{equation}
where \(M\) is the black hole mass, \(S\) is the Bekenstein–Hawking entropy, and \(\tau\) is the inverse temperature of the cavity enclosing the black hole, describing the off-shell thermodynamic state of the system. The on-shell condition is satisfied when \(\tau\) equals the Hawking inverse temperature  \(\beta=\frac{1}{T} \)\cite{a22}, at which the generalized free energy reduces to the standard Helmholtz free energy \(F=M-TS\). 
To characterize the topological structure of the thermodynamic parameter space, an auxiliary parameter \(\Theta \in (0, \pi)\) is introduced, and a modified free energy is defined as\cite{a18} 
\begin{equation}
    \tilde{\mathcal{F}} = \mathcal{F} + \frac{1}{\sin\Theta}.
\end{equation}
A two-component vector field is then constructed from the modified free energy\cite{a17,a18,a21,a22}
\begin{equation}
    \phi = \left( \phi^{r_+}, \phi^{\Theta} \right) = \left( \frac{\partial \tilde{\mathcal{F}}}{\partial r_+}, \frac{\partial \tilde{\mathcal{F}}}{\partial \Theta} \right),
\end{equation}
where \(r_+\) denotes the horizon radius of the black hole. The zeros of the radial component \(\phi^{rh}=0\) correspond to physically realizable black hole states in the thermodynamic parameter space, i.e., topological defects. The angular component \(\phi^{\Theta}\) diverges at the boundaries \(\Theta=0\) and \(\Theta=\pi\), ensuring the vector field points outward at the boundaries. 

Using the normalized vector field \(n^a = \phi^a / \|\phi\|\), a conserved topological current is constructed as\cite{a48,a55,a56} 
\begin{equation}
    j^{\mu} = \frac{1}{2\pi} \varepsilon^{\mu\nu\rho} \varepsilon_{ab} \partial_{\nu} n^{a} \partial_{\rho} n^{b}, \quad \mu, \nu, \rho = 0, 1, 2.
\end{equation}
This current satisfies the conservation law \(\partial_{\mu} j^{\mu} = 0\) and can be expressed in terms of the Jacobian determinant and the \(\delta \)-function\cite{a24,a25}
\begin{equation}
    j^{\mu} = \delta^{2}(\phi) J^{\mu}\left( \frac{\phi}{x} \right),
\end{equation}
\begin{table*}[htbp]
\centering
\caption{Thermodynamic properties of the black hole states for the nine topological (sub)classes of 
$W^{1-}$, $W^{0+}$, $W^{0-}$, $W^{1+}$, $W^{0-\leftrightarrow 1+}$, $\bar{W}^{1+}$, $\hat{W}^{1+}$, $\widetilde{W}^{1+}$, and $\ddot{W}^{1-}$, respectively.} 
\label{tab:xingzhi}   
\large
\renewcommand{\arraystretch}{1.3}
\resizebox{1\textwidth}{!}{%
\begin{tabular}{|c|c|c|c|c|c|c|}\hline

Topological (sub)classes & Innermost & Outermost & Low $T$ ($\beta\to\infty$) & High $T$ ($\beta\to0$) & DP & $W$ \\\hline

$W^{1-}$               & Unstable & Unstable & Unstable large & Unstable small & In pairs & $-1$ \\\hline
$W^{0+}$               & Stable   & Unstable & Unstable large + stable small & No & One more GP & $0$ \\\hline
$W^{0-}$               & Unstable & Stable   & No & Unstable small + stable large & One more AP & $0$ \\\hline
$W^{1+}$               & Stable   & Stable   & Stable small & Stable large & In pairs & $+1$ \\\hline
$W^{0-\leftrightarrow 1+}$ & Unstable & Stable & No & Stable large & One more AP & $0$ or $+1$ \\\hline
$\bar{W}^{1+}$         & Stable   & Stable   & No & Stable large & In pairs & $+1$ \\\hline
$\hat{W}^{1+}$         & Stable   & Stable   & Unstable small + two stable small & Stable large & One more GP & $+1$ \\\hline
$\widetilde{W}^{1+}$   & Unstable & Stable   & Stable small & Unstable small + stable small + stable large & One more AP & $+1$ \\\hline
$\ddot{W}^{1-}$& Unstable & Stable   & Unstable small & Unstable small + unstable small + stable large & One more AP & $-1$ \\ \hline

\end{tabular}%
}
\end{table*}
which is nonvanishing only at the zeros of the vector field.Integrating the time component of the topological current over the full parameter space yields the total topological number \(W\), which decomposes into the sum of local winding numbers of individual topological defects\cite{a18}
\begin{equation}
    W = \int_{\Sigma} j^{0} d^{2}x = \sum_{i=1}^{N} \beta_{i} \eta_{i} = \sum_{i=1}^{N} w_{i}.
\end{equation}
Here,\(w_i\) is the winding number for the \(i\)th zero point
of \(\phi\),  \(\beta _i\) is the Hopf index, counting the winding of the field around the zero point; \(\eta_i = \operatorname{sign}\left( J^0\left( \frac{\phi}{x} \right)_{z_i} \right)\) is the Brouwer degree, characterizing the orientation of the mapping. Their product defines the local winding number \(w_i\), an intrinsic invariant of the topological defect. A central corollary of this formalism establishes a direct correspondence between local winding numbers and the thermodynamic stability of black hole equilibria: configurations with \(w=+1\) are thermodynamically stable, while those bearing \(w=-1\) are thermodynamically unstable.

A complete classification scheme for black hole thermodynamic topology contains nine topological branches: \(W^{0+}\), \(W^{1-}\), \(W^{1+}\), \(W^{0-}\)\cite{a22}, \(\ddot{W}^{1-}\)\cite{a25}, \(\widetilde{W}^{1+}\)\cite{a24}, \(\hat{W}^{1+}\), \(\bar{W}^{1+}\), and \(W^{0-\leftrightarrow 1+}\)\cite{a23}. These branches are distinguished by stable configurations of inner or outer horizon solutions, arrangement of stable or unstable branches, phase profiles at low or high temperature limits, global topological numbers, and evolution of generation point (GP) and annihilation point (AP). This formalism holds for static\cite{a21,a22,a47}, rotating\cite{a57}, charged\cite{a58}, and gauged supergravity AdS black holes\cite{a59}.The nine branches are grouped into four fundamental topological families. The sign of heat capacity alternates monotonically with the  radius of outer horizon \(r_+\). Each family possesses invariant winding-number signs at the innermost and outermost horizons, and all intermediate equilibrium black hole solutions appear as conjugate \([+,-]\) pairs\cite{a22}. Table \ref{tab:xingzhi} summarizes branch stability, high/low-temperature thermodynamic properties, GP and AP behaviors and global topological charges for each family.

\section{TOPOLOGICAL ANALYSIS OF BARDEEN-ADS REGULAR BLACK HOLES AND DISCOVERY OF NEW SUBCLASS \(\widetilde{W}^{0-}\))}
\label{sec:level3}
The Bardeen–AdS-class black hole geometry emerges from the gravitational coupling of Einstein general relativity to nonlinear electrodynamics. The mass and magnetic charge of regular black holes are intrinsically correlated with the coupling constants of the matter sector, giving rise to the well-known coupling constant degeneracy. To resolve this unphysical constraint, a modified Lagrangian for nonlinear electromagnetic fields was constructed in Ref. \cite{a33}. 
\begin{equation}
\mathcal{L}_m = \frac{6m_0}{q_0^3} \left( \frac{\sqrt{q_0^2 F_{\mu\nu}F^{\mu\nu}/2}}{1+\sqrt{q_0^2 F_{\mu\nu}F^{\mu\nu}/2}} \right)^{5/2},
\end{equation}
Upon minimally coupling this Lagrangian to Einstein gravity, the full gravitational action reads
\begin{equation}\label{eq:8}
I=\tfrac{1}{16\pi}\!\int d^4x\sqrt{-g}\Bigg(R+\tfrac{6}{L^2}-\tfrac{6m_0}{q_0^3}
\times\left(\tfrac{\sqrt{q_0^2 F_{\mu\nu}F^{\mu\nu}/2}}{1+\sqrt{q_0^2 F_{\mu\nu}F^{\mu\nu}/2}}\right)^{\!5/2}\Bigg),
\end{equation}
where \(R\) is the Ricci scalar, \(L\) denotes the AdS radius, \(F_{\mu \nu }\) is the field-strength tensor of the nonlinear electromagnetic field, and \(m_0\), \(q_0\) are fixed coupling constants independent of the black hole state. This decoupling between coupling parameters and black hole physical quantities resolves the thermodynamic inconsistency inherent in the original Bardeen–AdS-class black hole. 
\(m_0\) and \(q_0\) serve simultaneously as interaction coupling constants and the ADM mass and magnetic charge of the black hole. To eliminate this inconsistency, we decouple their distinct physical roles: \(m_0\) and \(q_0\) are treated as fixed constants purely for couplings, while the physical mass and magnetic charge of the black hole are redefined via new parameters m and \(q_m\). Solving the field equations derived from Eq. \ref{eq:8}, we obtain the family of Bardeen–AdS-class black hole solutions.
\begin{equation}
    f(r) = \frac{r^2}{L^2} + 1 - \frac{m}{r} + m_0 \left( \frac{q_m}{q_0} \right)^{3/2} \left( \frac{1}{r} - \frac{r^2}{(r^2 + q_m q_0)^{3/2}} \right),
\end{equation}
where \(m\) is the mass parameter of the black hole, and \(q_m\) is the magnetic charge parameter. The event horizon radius \(r_+\) is determined by the horizon condition \(f(r_+) = 0\).The corresponding thermodynamic quantities follow directly from the horizon condition and fundamental thermodynamic relations, and are listed below 
\begin{align}
M &= \tfrac12m \nonumber \\
&= \tfrac12\Big( \tfrac{r_+^3}{L^2} + r_+ + m_0\big(\tfrac{q_m}{q_0}\big)^{3/2}
\Big( 1 - \tfrac{r_+^3}{\big(r_+^2 + q_0 q_m\big)^{3/2}} \Big) \Big),
\label{eq:10}
\end{align}
\begin{equation}
    S = \pi r_+^2.
    \label{eq:11}
\end{equation}
By inserting eq. \ref{eq:10} and eq. \ref{eq:11} into eq. \ref{eq:1}, we can derive
\begin{equation}
\begin{aligned}
\small
\mathcal{F} &= M - \tfrac{S}{\tau} \\
&= -\tfrac{\pi r_+^2}{\tau}+\tfrac12\Big(r_++\tfrac{r_+^3}{L^2}+m_0\big(\tfrac{q_m}{q_0}\big)^{3/2}\big(1-\tfrac{r_+^3}{(r_+^2+q_0q_m)^{3/2}}\big)\Big),
\end{aligned}
\end{equation}

as a result, the elements of the vector \(\phi \) are
\begin{equation}
\phi ^{r_+}= \frac{1}{2}\left(
1 + \frac{3 r_+^2}{L^2} - \frac{4 \pi r_+}{\tau}
- \frac{3 r_+^2 m_0 q_m^2 \sqrt{\dfrac{q_m}{q_0}}}{\left(r_+^2 + q_0 q_m\right)^{5/2}}
\right),
\end{equation}
\begin{equation}
    \phi^{\Theta} = -\cot \Theta \csc \Theta.
\end{equation}
taking the boundary condition \(\phi^{r_+}
= 0\), the inverse temperature parameter \(\tau\) can be expressed as
\begin{equation}
\tau = \beta = \frac{4\pi r_+}{1 + \dfrac{3 r_+^2}{L^2} - \dfrac{3 r_+^2 m_0 q_m^2 \sqrt{\dfrac{q_m}{q_0}}}{\left(r_+^2 + q_0 q_m\right)^{5/2}}}.
\end{equation}

A critical quantity \(\hat{m}_{01}/q_0\) and \(\hat{m}_{02}/q_0\) was defined in Ref.\cite{a38} to characterize the parameter space of Bardeen–AdS-class black holes, whose explicit expression reads
\begin{equation}
    \frac{\hat{m}_{01}}{q_0} = \frac{\sqrt{6} \left( 6 - L^2/q_0^2 + \sqrt{(L^2/q_0^2)^2 + 24L^2/q_0^2} \right)^{5/2}}{9 \left( -L^2/q_0^2 + \sqrt{(L^2/q_0^2)^2 + 24L^2/q_0^2} \right)^2}.
\end{equation}
\begin{equation}
\frac{\hat{m}_{02}}{q_0} = \frac{\left(30 - 3L^2/q_0^2 + \sqrt{3L^2/q_0^2\left(3L^2/q_0^2 + 40\right)}\right)^{7/2}}{225\sqrt{30}\left(-3L^2/q_0^2 + \sqrt{3L^2/q_0^2\left(3L^2/q_0^2 + 40\right)}\right)^2},
\label{eq:17}
\end{equation}

The critical parameter \(\hat{m}_{01}\) separates Type I and Type II Bardeen–AdS-class black holes. For fixed \(L\) and \(q_0\), configurations with \(m_0\geq\hat{m}_{01}\) belong to Type I, while those satisfying \(0<m_0<\hat{m}_{01}\) fall into the Type II regime. Within the Type II parameter space, an additional critical value \(\hat{m}_{02}\) provides a finer subdivision: for \(\hat{m}_{02}<m_0<\hat{m}_{01}\), the radius of the outer horizon \(r_+\) occupies two disconnected domains, giving rise to discontinuities in the thermodynamic characteristic curves originating from multiple horizons; on the contrary, for \(0<m_0\leq\hat{m}_{02}\), the admissible range of \(r_+\) forms a single continuous interval, and the thermodynamic curves remain fully continuous everywhere.

Prior to the numerical topological analysis for Type I and Type II Bardeen–AdS-class black holes, we analytically investigated the asymptotic behaviors of the inverse Hawking temperature \(\beta(r_{+})\) at two extreme horizon limits: \(r_{+}\to0\) and \(r_{+}\to\infty\) (infinitely large horizon). This dual asymptotic signature serves as the pivotal criterion for identifying the topological family \(W^{0-}\). Throughout this work, we adopt dimensionless simplification \(q_{0}=q_{m}=1\), and the corresponding limiting expressions are presented below
\begin{equation}
\lim_{r_+ \to r_m} \beta(r_+) = 0, \lim_{r_+ \to \infty} \beta(r_+) = 0.
\label{eq:18}
\end{equation}

Here \(r_m\) denotes the minimal physically accessible horizon radius, which can either vanish or take a finite value. For fixed-charge RN black holes, the extremal horizon yields \(r_m=M=Q=r_e\), while \(r_m=0\) holds for Schwarzschild spacetime.
\subsection{\label{sec:leve3A} Topological Analysis of Type I Black Holes}
We perform comparative numerical calculations with two AdS curvature radii \(L=1\) and \(L=15\) to obtain the critical threshold \(\hat{m}_{01}\) that distinguishes Type I and Type II black holes. Specifically, \(\hat{m}_{01}\approx5.37914\) for \(L=1\) and \(\hat{m}_{01}\approx2.62087\) for \(L=15\). According to the classification criteria proposed in previous sections, five sets of coupling constants \(m_0\) are examined for each curvature parameter: \(m_0=8,\;5.37914,\;4.06032,\;5,\;3\) for \(L=1\) and \(m_0=3,\;2.62087,\;1.80887,\;2,\;1.6\) for \(L=15\). For both parameter groups, the first two cases with \(m_0\ge\hat{m}_{01}\) correspond to Type I black hole geometries, while the remaining three cases with \(m_0<\hat{m}_{01}\) belong to Type II black holes. Figures \ref{fig:L=1,m_0=8Type_I} and \ref{fig:L=15,m_0=3Type_I} present the evolutionary curves of the horizon radius \(r_+\) with inverse temperature \(\beta=1/T\) for two representative Type I parameter configurations.
\begin{figure}[htbp]
  \centering
  \includegraphics[width=0.35\textwidth]{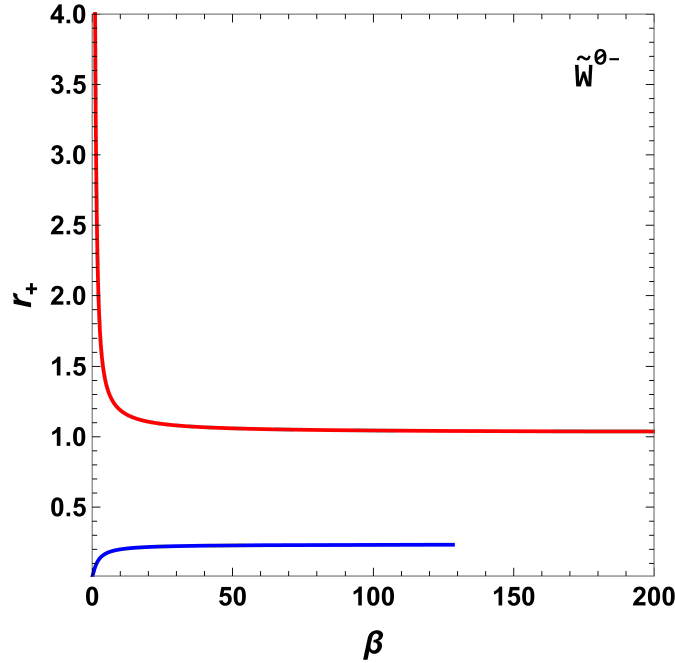}
  \caption{Vector field \(\phi^{r_+}\) zero points on the \(r_+\)–\(\beta\) plane for the Bardeen-AdS black hole with \(L=1, m_0=8\). Blue: unstable branch (\(w=-1\)); red: stable branch (\(w=+1\)).}
  \label{fig:L=1,m_0=8Type_I} 
\end{figure}

Figure \ref{fig:L=1,m_0=8Type_I} corresponds to the configuration with \(L=1\) and \(m_0=8\). The thermodynamic curve contains one stable solution branch and one unstable solution branch, yielding a total topological winding number \(W=-1+1=0\), and the winding number sequence of the inner and outer boundary branches is \([-,+]\). The asymptotic boundary behavior observed from the numerical curves is consistent with the analytical results of Eq. \ref{eq:18}:
\(\beta (r_m)=0,\quad \beta (\infty)=0\).

The topological features of this configuration are similar to the \(W^{0-}\) topological class listed in Tab. \ref{tab:xingzhi}, with distinct evolutionary behaviors. Thermodynamically, only stable small black holes exist at low temperatures, while unstable small black holes coexist with stable large black holes at high temperatures, and no topological GP or AP emerge on the thermodynamic curve. The numerical profile consists of two isolated curve branches. Topologically, this configuration evolves from an unstable background state with the winding number signature \([-,-]\), and finally forms the \([-,+]\) winding sequence by appending one new stable branch labeled \([+]\) at the end of the original sequence. As demonstrated in Ref. \cite{a22}, the conventional \(W^{0-}\) topology realizes topological expansion by inserting internal \([+,-]\) winding pairs inside the sequence. In contrast, the present configuration produces no internal variation of winding pairs and achieves topological evolution merely via terminal branch extension. For this distinct topological evolution mechanism, we define this configuration as a novel secondary subclass \(\widetilde{W}^{0-}\) that resides exclusively within the pre-existing \(W^{0-}\) topological family.
\begin{figure}[htbp]
  \centering
  \includegraphics[width=0.35\textwidth]{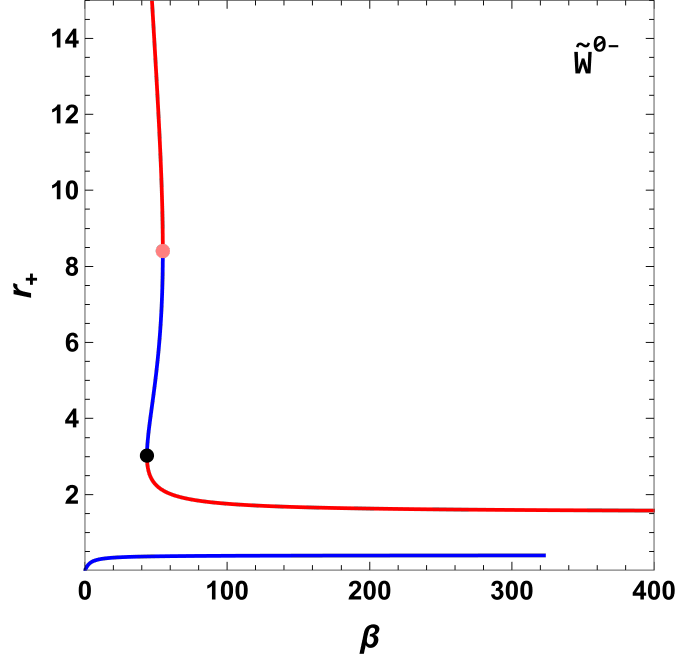}
  \caption{Vector field \(\phi^{r_+}\) zero points on the \(r_+\)–\(\beta\) plane for the Bardeen-AdS black hole with \(L=15, m_0=3\). Blue: unstable branch (\(w=-1\)); red: stable branch (\(w=+1\)). Pink dot denote AP, and black dot denote GP.}
  \label{fig:L=15,m_0=3Type_I} 
\end{figure}
\begin{figure}[htbp]
  \centering
  \includegraphics[width=0.35\textwidth]{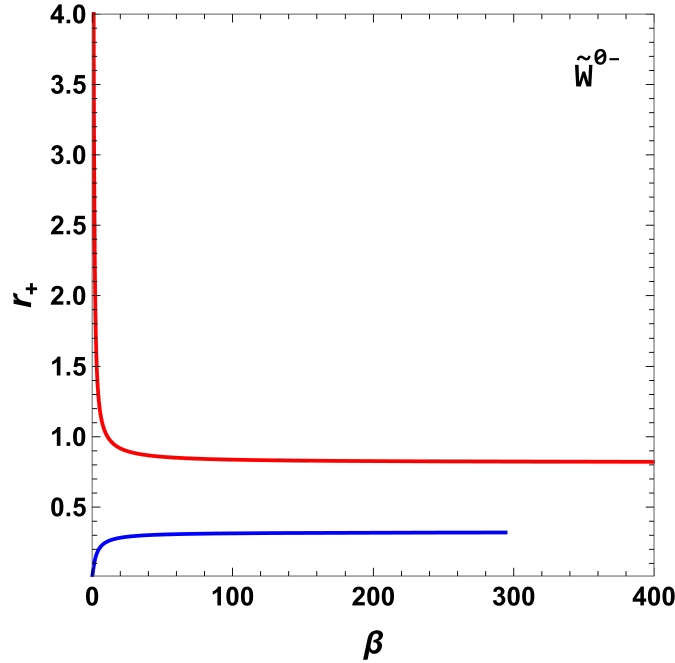}
  \caption{Vector field \(\phi^{r_+}\) zero points on the \(r_+\)–\(\beta\) plane for the Bardeen-AdS black hole with \(L=1, m_0=5.37914\). Blue: unstable branch (\(w=-1\)); red: stable branch (\(w=+1\)).}
  \label{fig:L=1,m_0=5.37914Type_I} 
\end{figure}
\begin{figure}[htbp]
  \centering
  \includegraphics[width=0.35\textwidth]{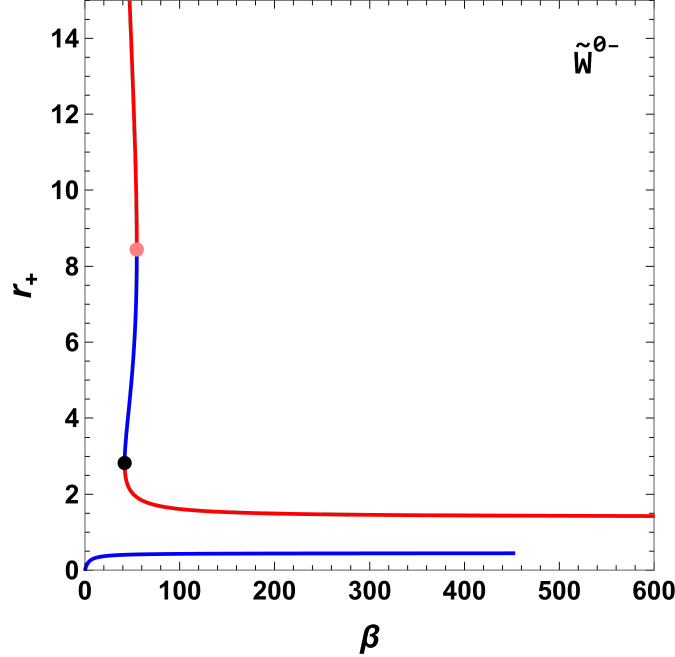}
  \caption{Vector field \(\phi^{r_+}\) zero points on the \(r_+\)–\(\beta\) plane for the Bardeen-AdS black hole with \(L=15, m_0=2.62087\). Blue: unstable branch (\(w=-1\)); red: stable branch (\(w=+1\)). Pink dot denote AP, and black dot denote GP.}
  \label{fig:L=15,m_0=2.62087Type_I} 
\end{figure}

Figure \ref{fig:L=15,m_0=3Type_I} plots the results for the configuration \(L=15,\,m_0=3\). The corresponding thermodynamic curve includes two stable and two unstable branches and satisfies the asymptotic relations \(\beta (r_m)=0,\;\beta (\infty)=0\) given in Eq. \ref{eq:18}. The total topological winding number is calculated as \(W=-1+1-1+1=0\). While the inner and outer boundary branches retain the fundamental winding signature \([-,+]\), an additional \([+,-]\) winding pair arises in the middle of the curve, accompanied by one generation point and one annihilation point. Such topological expansion via embedded internal winding pairs belongs to the unified topological framework, which confirms that this configuration is also a member of the \(\widetilde{W}^{0-}\) subclass. The emergence of paired stable and unstable branches significantly enriches the thermodynamic solution spectrum. Specifically, only stable small black holes survive at low temperatures, while four types of black hole solutions, namely unstable small, stable small, unstable intermediate, and stable large black holes, coexist at high temperatures.

Figures \ref{fig:L=1,m_0=5.37914Type_I} and \ref{fig:L=15,m_0=2.62087Type_I} display the \(r_+-\beta\) evolutionary curves for the critical configurations \(L=1,\,m_0=5.37914\) and \(L=15,\,m_0=2.62087\), respectively. The topological profile, branch structure and winding number evolution behavior of the two critical configurations are entirely consistent with those of the aforementioned typical Type I black hole configurations, which further verifies that the critical systems also belong to the \(\widetilde{W}^{0-}\) topological subclass.
\subsection{\label{sec:leve3B} Topological Investigation of Type II Black Holes}
This chapter systematically investigates the topological properties of Type II Bardeen–AdS-class black holes. Figures \ref{fig:L=1,m_0=4.06032Type_II} and \ref{fig:L=15,m_0=1.80887Type_II} display the thermodynamic evolution of the event horizon radius \(r_+\) as a function of the inverse temperature \(\beta=1/T\) for Bardeen–AdS-class black holes with parameter sets \((L=1,\,m_0=4.06032)\) and \((L=15,\,m_0=1.80887)\), respectively. The global topological profiles of the two configurations are highly consistent with those presented in Figs. \ref{fig:L=1,m_0=8Type_I} and \ref{fig:L=15,m_0=3Type_I}, while their thermodynamic curves exhibit evident morphological distinctions. For coupling parameters satisfying \(m_0\geq\hat{m}_{01}\), the lower unstable blue thermodynamic branch is confined to a finite parameter interval with clear boundaries on both sides. By contrast, when the coupling constant takes the secondary critical value \(m_0=\hat{m}_{02}\), this unstable branch can continuously extend toward the low-temperature limit \(\beta\to\infty\). 
\begin{figure}[htbp]
  \centering
  \includegraphics[width=0.35\textwidth]{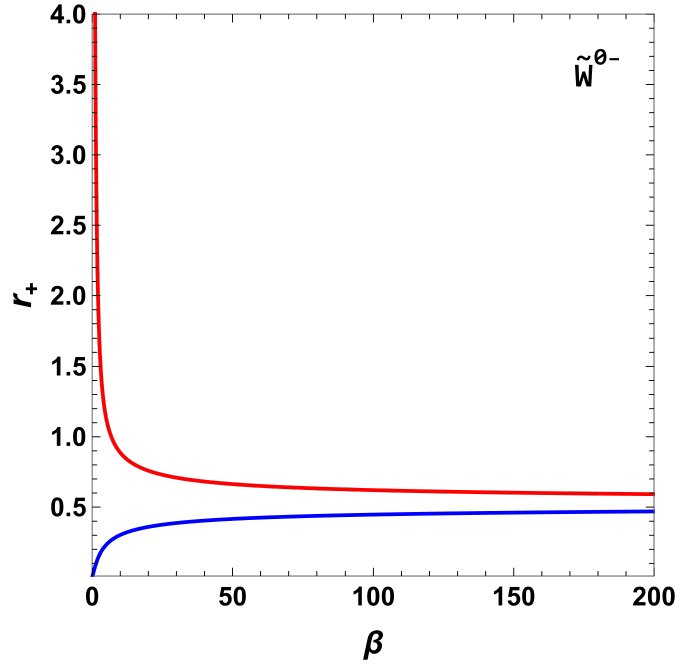}
  \caption{Vector field \(\phi^{r_+}\) zero points on the \(r_+\)–\(\beta\) plane for the Bardeen-AdS black hole with \(L=1, m_0=4.06032\). Blue: unstable branch (\(w=-1\)); red: stable branch (\(w=+1\)).}
  \label{fig:L=1,m_0=4.06032Type_II} 
\end{figure}
\begin{figure}[htbp]
  \centering
  \includegraphics[width=0.35\textwidth]{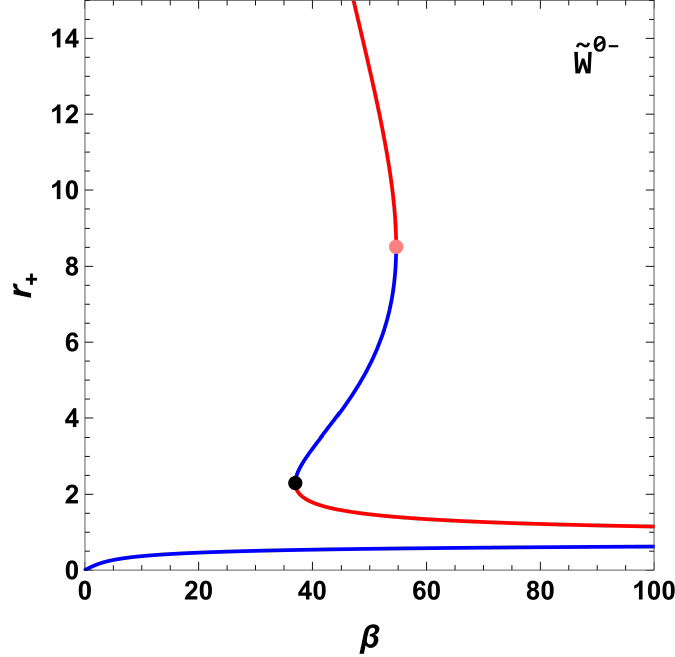}
  \caption{Vector field \(\phi^{r_+}\) zero points on the \(r_+\)–\(\beta\) plane for the Bardeen-AdS black hole with \(L=15, m_0=1.80887\). Blue: unstable branch (\(w=-1\)); red: stable branch (\(w=+1\)). Pink dot denote AP, and black dot denote GP.}
  \label{fig:L=15,m_0=1.80887Type_II} 
\end{figure}

Despite such differences in curve extension behavior, the two configurations share identical core topological criteria: the winding-number sequence of the inner and outer branches is \([-,+]\), the asymptotic behaviors satisfy \(\beta(r_m)=0\) and \(\beta(\infty)=0\), the global topological number reads \(W=-1+1=0\), and all curves possess piecewise discontinuous structures containing exactly one stable branch and one unstable branch. Consequently, both configurations are classified as the novel topological subclass \(\widetilde{W}^{0-}\) proposed in this work.
\begin{figure}[htbp]
  \centering
  \includegraphics[width=0.35\textwidth]{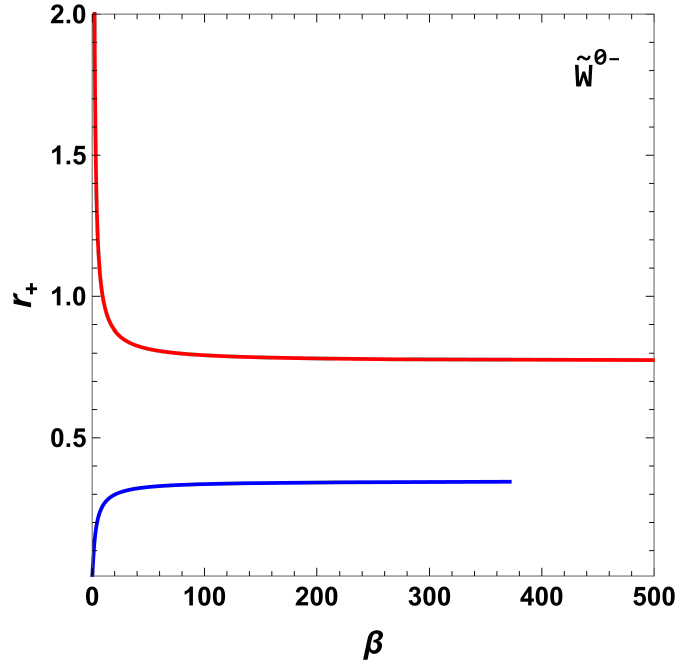}
  \caption{Vector field \(\phi^{r_+}\) zero points on the \(r_+\)–\(\beta\) plane for the Bardeen-AdS black hole with \(L=1, m_0=5\). Blue: unstable branch (\(w=-1\)); red: stable branch (\(w=+1\)).}
  \label{fig:L=1,m_0=5Type_II} 
\end{figure}
\begin{figure}[htbp]
  \centering
  \includegraphics[width=0.35\textwidth]{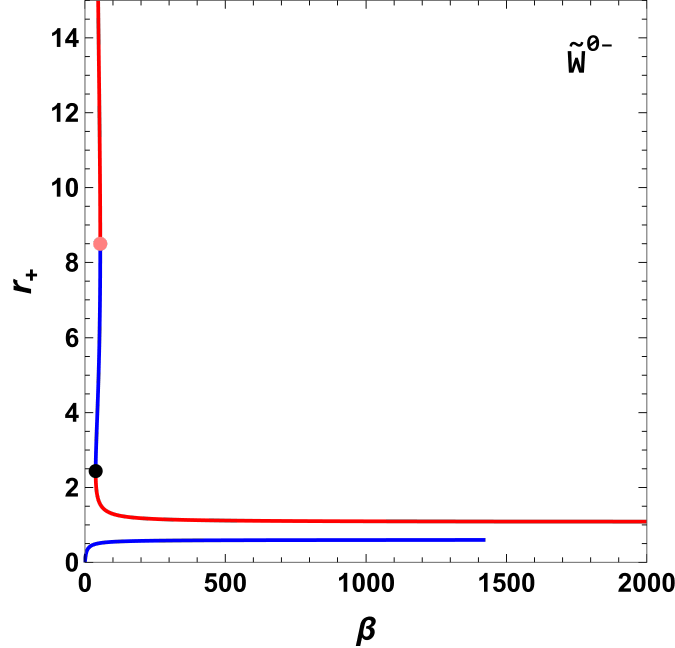}
  \caption{Vector field \(\phi^{r_+}\) zero points on the \(r_+\)–\(\beta\) plane for the Bardeen-AdS black hole with \(L=15, m_0=2\). Blue: unstable branch (\(w=-1\)); red: stable branch (\(w=+1\)). Pink dot denote AP, and black dot denote GP. }
  \label{fig:L=15,m_0=2Type_II} 
\end{figure}

Figures \ref{fig:L=1,m_0=5Type_II} and \ref{fig:L=15,m_0=2Type_II} further illustrate the variation of \(r_+\) with \(\beta\) for Bardeen–AdS-class black holes with parameters \(L=1,\,m_0=5\) and \(L=15,\,m_0=2\). The topological structures of the two cases perfectly match the I-type benchmark results shown above, which further verifies that their topological features satisfy the classification criteria of the \(\widetilde{W}^{0-}\) subclass.
\begin{figure}[htbp]
  \centering
  \includegraphics[width=0.35\textwidth]{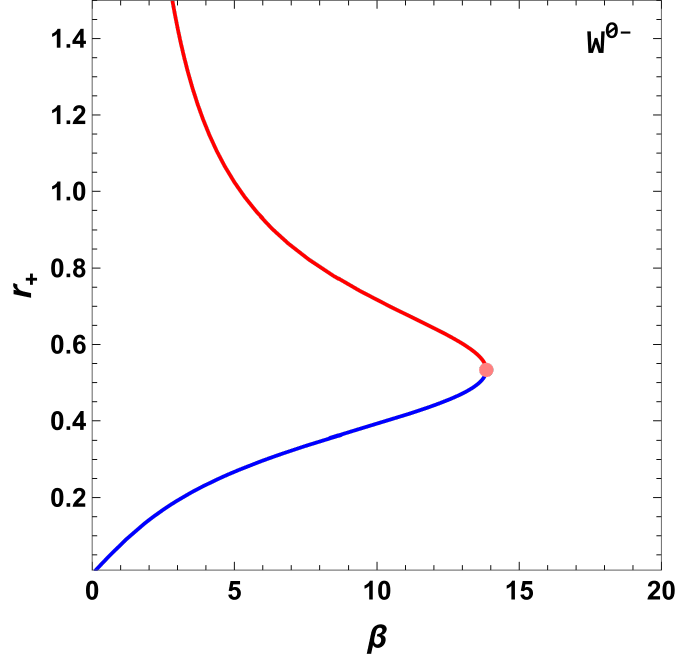}
  \caption{Vector field \(\phi^{r_+}\) zero points on the \(r_+\)–\(\beta\) plane for the Bardeen-AdS black hole with \(L=1, m_0=3\). Blue: unstable branch (\(w=-1\)); red: stable branch (\(w=+1\)). Pink dot denote AP.}
  \label{fig:L=1,m_0=3Type_II} 
\end{figure}
\begin{figure}[htbp]
  \centering
  \includegraphics[width=0.35\textwidth]{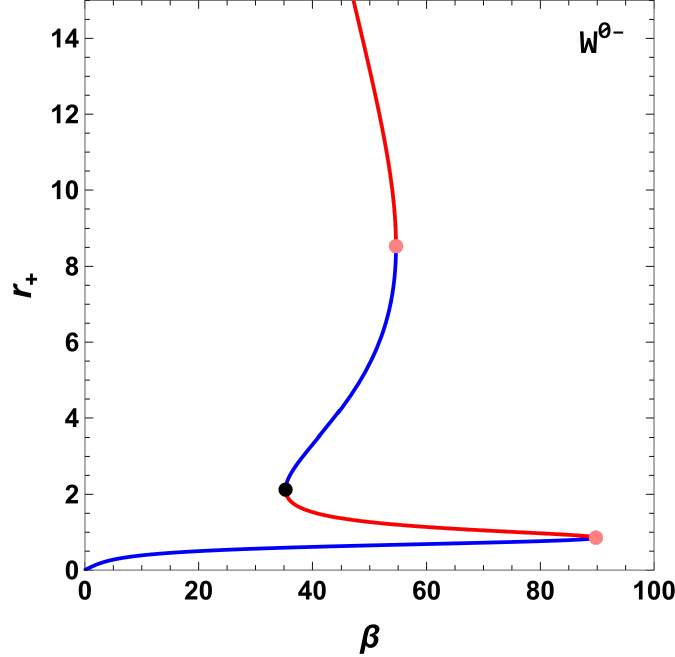}
  \caption{Vector field \(\phi^{r_+}\) zero points on the \(r_+\)–\(\beta\) plane for the Bardeen-AdS black hole with \(L=15, m_0=1.6\). Blue: unstable branch (\(w=-1\)); red: stable branch (\(w=+1\)). Pink dots denote APs, and black dot denote GP.}
  \label{fig:L=15,m_0=1.6Type_II} 
\end{figure}

Finally, Figs. \ref{fig:L=1,m_0=3Type_II} and \ref{fig:L=15,m_0=1.6Type_II} plot the complete thermodynamic evolutionary curves for parameter combinations \((L=1,\,m_0=3)\) and \((L=15,\,m_0=1.6)\). Both configurations yield the same global topological number \(W=0\), the asymptotic boundary behavior observed from the numerical curves is consistent with the analytical results of Eq. \ref{eq:18}: \(\beta (r_m)=0,\quad \beta (\infty)=0\).

For the curve corresponding to \((L=1,\,m_0=3)\), no valid black hole solutions exist in the low-temperature regime, while unstable small black holes and stable large black holes coexist at the high-temperature limit. The winding-number combination of the innermost and outermost branches takes the form \([-,+]\), which belongs to the standard \(W^{0-}\) topological family. In contrast, the curve for \((L=15,\,m_0=1.6)\) introduces an additional paired winding-number unit \([+,-]\) in the middle of the fundamental \([-,+]\) sequence. Following the topological classification rules established previously, extended configurations with embedded paired winding-number structures within the base sequence still fall into the same topological framework, so this configuration is also categorized as the \(W^{0-}\) class.

For a concise overview of topological phase structures across the full parameter space, we tabulate all key thermodynamic and topological quantities in Table \ref{tab:2}.

\begin{table*}[htbp]
\centering
\caption{Summary of topological and thermodynamic properties for Type I and Type II Bardeen–AdS-class black holes with two AdS radius $L=1$ and $L=15$}
\label{tab:2}
\renewcommand{\arraystretch}{1.3}

\begin{tabular}{|c|c|c|c|c|c|c|}
\hline
AdS radius& Black hole type& r range& DP&Low $T$ ($\beta\to\infty$)  &High $T$ ($\beta\to0$) & class\\
\hline
\multirow{5}{*}{$L=1$} & \multirow{2}{*}{Type I} & $m_0>\hat{m}_{01}$ & 0 & $S_s$& $U_s$+$S_l$& $\widetilde{W}^{0-}$ \\
\cline{3-7}
& & $m_0=\hat{m}_{01}$ & 0 & $S_s$& $U_s$+$S_l$& $\widetilde{W}^{0-}$ \\
\cline{2-7}
& \multirow{3}{*}{Type II} & $m_0=\hat{m}_{02}$ & 0 & $S_s$+$U_s$& $U_s$+$S_l$& $\widetilde{W}^{0-}$ \\
\cline{3-7}
& & $\hat{m}_{02}<m_0<\hat{m}_{01}$ & 0 & $S_s$& $U_s$+$S_l$& $\widetilde{W}^{0-}$ \\
\cline{3-7}
& & $m_0<\hat{m}_{02}$ & One more AP & No & $U_s$+$S_l$& $W^{0-}$ \\
\hline
\multirow{5}{*}{$L=15$} & \multirow{2}{*}{Type I} & $m_0>\hat{m}_{01}$ & In pairs & $S_s$& $U_s$+$S_s$+$U_m$+$S_l$& $\widetilde{W}^{0-}$ \\
\cline{3-7}
& & $m_0=\hat{m}_{01}$ & In pairs & $S_s$& $U_s$+$S_s$+$U_m$+$S_l$& $\widetilde{W}^{0-}$ \\
\cline{2-7}
& \multirow{3}{*}{Type II} & $m_0=\hat{m}_{02}$ & In pairs & $S_s$+$U_s$& $U_s$+$S_s$+$U_m$+$S_l$& $\widetilde{W}^{0-}$ \\
\cline{3-7}
& & $\hat{m}_{02}<m_0<\hat{m}_{01}$ & In pairs & $S_s$& $U_s$+$S_s$+$U_m$+$S_l$& $\widetilde{W}^{0-}$ \\
\cline{3-7}
& & $m_0<\hat{m}_{02}$ & One more AP & No & $U_s$+$S_s$+$U_m$+$S_l$& $W^{0-}$ \\
\hline
\end{tabular}

\begin{tablenotes}
\item \(U\) denotes unstable states and \(S\) denotes stable states; the subscripts \(s\), \(m\), \(l\) correspond to small, intermediate and large black holes, respectively.
\end{tablenotes}
\end{table*}
\begin{table*}[htbp]
\centering
\caption{Comparison of thermodynamic and topological properties between the novel subclass \(\widetilde{W}^{0-}\) and the canonical \(W^{0-}\) topological family}
\label{tab:newclass}
\large
\renewcommand{\arraystretch}{1.3}
\resizebox{1\textwidth}{!}{%
\begin{tabular}{|c|c|c|c|c|c|c|}\hline
Topological (sub)classes & Innermost & Outermost & Low \(T\) (\(\beta\to\infty\)) & High \(T\) (\(\beta\to0\)) & DP & \(W\) \\\hline
\(\widetilde{W}^{0-}\)& Unstable  & Stable    & stable small& Unstable small + stable large  & No& \(0\) \\\hline
 $W^{0-}$               & Unstable & Stable   & No & Unstable small + stable large & One more AP &$0$ \\\hline
\end{tabular}%
}
\end{table*}
\section{Conclusions}
\label{sec:level4}
In this work, we investigate regular Bardeen–AdS-class black holes free of central curvature singularities. Adopting the thermodynamic topological formalism built upon the \(\varphi \)-mapping topological current and winding number invariants, we treat black hole equilibrium solutions as topological defects in the thermodynamic phase space and systematically perform a universal topological classification for such singularity-free spacetimes. We carry out comprehensive numerical topological calculations over multiple matter coupling parameters with two representative AdS curvature radii \(L=1\) and \(L=15\). We identify a novel topological subclass \(\widetilde{W}^{0-}\) within the conventional \(W^{0-}\) topological family, which extends the established nine-category classification framework for black hole thermodynamics.

All Bardeen–AdS-class black holes fall into the \(W^{0-}\) topological family characterized by the winding-number sequence \([-,+]\) and obey a unified set of topological signatures: the winding-number arrangement for the innermost and outermost equilibrium branches reads \([-,+]\), the global total topological number identically satisfies \(W=0\), and the inverse Hawking temperature asymptotically vanishes in both low- and high-temperature limits. Nevertheless, the subclass\(\widetilde{W}^{0-}\) possesses a distinctive topological evolution mechanism that distinguishes it from canonical \(W^{0-}\) configurations. Whereas the conventional \(W^{0-}\) topology expands by inserting paired \((+,-)\) winding-number units inside the base sequence, \(\widetilde{W}^{0-}\) evolves solely by appending a single stable branch at the terminus of the primitive unstable sequence without generating any internal paired winding-number structures.

The phase space of regular Bardeen–AdS-class black holes is split into Type I and Type II regimes by the critical coupling constants \(\hat{m}_{01}\) and \(\hat{m}_{02}\). All Type I black holes with \(m_0\ge\hat{m}_{01}\) belong uniformly to the \(\widetilde{W}^{0-}\) subclass. Their thermodynamic curves contain one stable and one unstable equilibrium branch with no GP or AP. Only stable small black holes populate the low-temperature regime, while unstable small black holes coexist with stable large black holes at high temperatures. At the critical threshold \(m_0=\hat{m}_{01}\), the system develops a four-branch phase structure accompanied by paired GP and AP, where four thermodynamic states—stable large, unstable intermediate, stable small, and unstable small black holes—coexist at high temperatures. Even so, its core winding-number sequence and asymptotic topological properties still satisfy the classification criteria for \(\widetilde{W}^{0-}\). Table \ref{tab:newclass} summarizes the core properties of the new subclass \(\widetilde{W}^{0-}\) and illustrates its distinctions from the original \(W^{0-}\) family.

Type II black holes correspond to the parameter range \(0<m_0<\hat{m}_{01}\), which is further partitioned by the secondary critical parameter \(\hat{m}_{02}\). For \(\hat{m}_{02}<m_0<\hat{m}_{01}\), the thermodynamic curves exhibit piecewise discontinuities with a truncated domain for the unstable branch, yet the spacetime remains classified as \(\widetilde{W}^{0-}\). In contrast, the horizon domain becomes continuous for \(0<m_0\le\hat{m}_{02}\), and the system reduces to the standard \(W^{0-}\) topological configuration.

Numerical calculations covering the full parameter space confirm the self-consistency of the thermodynamic topological method for the nonlinear-electrodynamics-coupled regular Bardeen–AdS-class black hole model studied in this work. We refine the topological classification criteria specific to this family of singularity-free black holes and uncover their distinctive topological evolution and multiphase coexistence behaviors, demonstrating that the topological framework can be extended to nonsingular spacetimes. Our results provide a reference for topological investigations of higher-dimensional, rotating, multi-charged regular black holes within the same theoretical setup, and further calculations on additional singularity-free gravitational models are required to test the generality of the identified topological rules.

\begin{acknowledgments}

This work was supported by the National Natural Science Foundation of China (No.12265007), and the Guizhou Provincial Major Scientific and Technological Program (XKBF(2025)010).
\end{acknowledgments}

\bibliographystyle{unsrt}
\bibliography{wenxian}
\end{document}